# Range resolution improvement for two close targets by using the FM signals

R. Ghavamirad, R.A. Sadeghzadeh[*], M.A. Sebt, A.H. Naderi

In this paper, we have focused on the improvement of the range resolution of two close targets by using FM signal. Two algorithms comprise IFFT and MUSIC have been used to reach our goals. On both algorithms, we have tried to improve the range resolution by using more FM channels. The simulation results show that the music method by using the same FM channel gives us a good range resolution twice as well as IFFT method.

*Introduction:* Echoes that are usually received from antennas will be amplified, and after that, their frequency will suppress. Afterward, it passes from a threshold comparator circuit in which the signal envelope extraction is done. The signal's envelope is proportional to power of received reflections [1], [2].

The Range resolution of two targets in constant false alarm radar systems is inversely proportional to the transmitted signal bandwidth [3], [4]. According to this, the range resolution of close targets could be improved by increasing modulation bandwidth in FM broadcasting radar systems [5-7].

The modulation bandwidth could be increased by using more FM channels in FM broadcasting systems. A maximum bandwidth of 20 MHz could be employed in the spectrum of FM broadcasting radio which occupies the frequency band of 88 to 108 MHz [4].

In this paper, three states of one, three and seven FM channel broadcasting, has been investigated. Each of these states examined by two detections methods including inverse Fourier transforms (IFFT) and multiple signal classification (MUSIC) algorithm.
Two targets with the distance of 3 km from each other has been detected easily by IFFT detection method while in this case the
 two targets at a distance of 2 km detected easily by applying MUSIC detection method .with the same procedure the IFFT method applied to 3 and 7 FM channel signal for detecting two targets at a distance of 2 and 1 km, respectively .with the ever employing MUSIC method the range resolution is improved about 1 km, that means For 3 and 7 FM channels signal the two targets has been detected easily with a distance of 1km and 300m, respectively. Within these results, the MUSIC method compared to IFFT method in the corresponding state of occupied FM channels, improved the range resolution about 1km. comparatively, the range resolution of two close targets by applying MUSIC algorithm is twice as good as IFFT method.

In section 2 the detection based on IFFT is reviewed then the multiple signal classification algorithm has discussed in chapter 3. The simulation results are depicted in chapter 4 and finally conclusions are described in chapter 5.

*Detection based on IFFT*: To detect targets, the received signal is first transmitted to the receiver and then its frequency would suppress to intermediate frequency (IF) in mixer. After that the signal passes through analog-to-digital converter (ADC) to be processed as a digital signal. Fig. 1 shows the detector's block diagram based on the direct signal and the signal received from the receiver's antenna which is depicted in detail in Fig. 2.

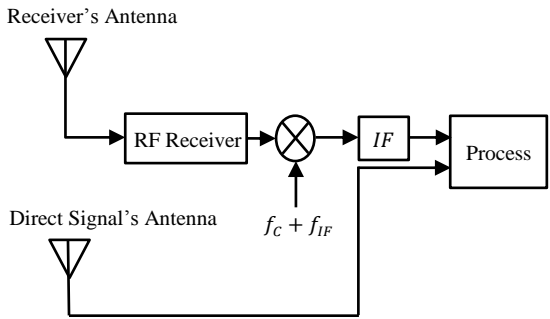

**Fig. 1** *Detection diagram*

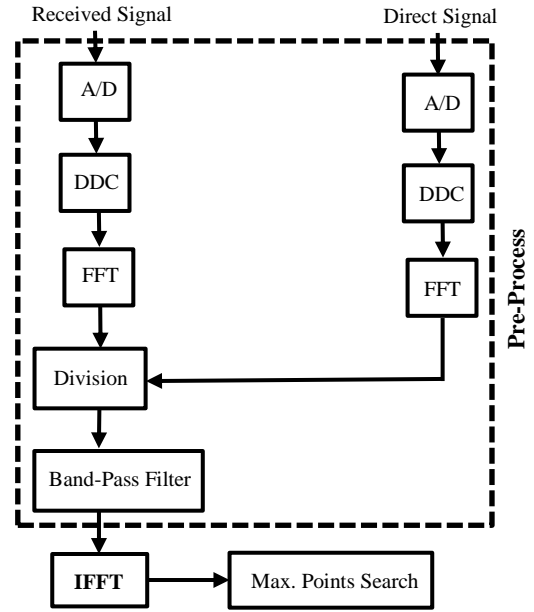

**Fig. 2** *Process box based on IFFT Method*

In this method, the direct signal and received antenna's signal are sampled by passing through ADC and then its frequency will decrease by digital down converter (DDC) in order to perform calculations in baseband. Afterward, the calculations of Fourier transform of the reference signal and antenna's received signal is performed. Then the results of Fourier transform of antenna's received signal divide by direct signal Fourier transform is obtained easily. This term includes the delay of target's received signal. By crossing this term through band pass filter (which it's bandwidth is relatively equal to received signal bandwidth) and performing IFFT a new term is obtained that its real part has some maximum points which are desirable.

Consider $s(t)$ and $s_r(t)$ as direct signal and received target signal in receiver's antenna respectively.

$$s(t) = A \exp\left(j2\pi\left(f_c t + f_\Delta \int_0^t x_m(\lambda)d\lambda\right)\right) \quad (1)$$

$$s_r(t) = A_r \exp\left(j2\pi\left(f_c(t - t_0) + f_\Delta \int_0^{t-t_0} x_m(\lambda)d\lambda\right)\right) \quad (2)$$

In (1) and (2), $A$ and $A_r$ are the amplitude of the direct signal and received target signal, respectively. $f_c$ is the carrier frequency and $f_\Delta$ is the maximum frequency deviation. $x_m(\lambda)$ is the direct message signal. According to detection algorithm the antenna's received signal should pass from RF receivers and enters the mixer block. After sampling and mitigating its frequency by DDC block, its FFT would calculated by

$$S_r(f) = \int_{-\infty}^{+\infty} s_r(t') \exp(-j2\pi f t') dt'$$

$$= A_r \int_{-\infty}^{+\infty} \exp\left(j2\pi\left((f_c - f)t' - f_c t_0 + f_\Delta \int_0^{t-t_0} x_m(\lambda)d\lambda\right)\right) dt' \quad (3)$$

Also these procedures are performed on direct message signal:

$$S(f) = \int_{-\infty}^{+\infty} s(t') \exp(-j2\pi f t') dt'$$

$$= A \int_{-\infty}^{+\infty} \exp\left(j2\pi\left((f_c - f)t' + f_\Delta \int_0^{t'} x_m(\lambda)d\lambda\right)\right) dt' \quad (4)$$

Then division operation which is shown in Fig. 2 should be performed.

$$D(f) = S_r(f)/S(f) \quad (5)$$

Crossing $D(f)$ from band-pass filter and calculating its IFFT and probing its maximum points, would obtained the range resolution of targets finally.

*Detection basis on MUSIC algorithm*: Multiple signal classification algorithm is one of the newest algorithm of detecting the direction of targets with a high range resolution [8], [9]. Consider $x[n]$ as desired signal.

$$x[n] = a_0 e^{j\omega_0 n} + a_1 e^{j\omega_1 n} + w[n] \quad (6)$$
$$n = 0, 1, \ldots, N-1$$

where $a_0$ and $a_1$ are signal's attenuation coefficient and $\omega_0$ and $\omega_1$ are angular frequency of signal. $w[n]$ is additive Gaussian White Noise signal with average of zero and variance of $\sigma^2$.
define **x** as:

$$\underline{x} = \begin{bmatrix} x[0] \\ \vdots \\ x[N-1] \end{bmatrix} \quad (7)$$

if , **a** and **w** define as follows, by substituting these into 7, **x** obtained.

$$\mathbf{S} = \begin{bmatrix} 1 & 1 \\ e^{j\omega_0} & e^{j\omega_1} \\ \vdots & \vdots \\ e^{j\omega_0(N-1)} & e^{j\omega_1(N-1)} \end{bmatrix} \quad (8)$$

$$\underline{\mathbf{a}} = \begin{bmatrix} a_1 \\ a_2 \end{bmatrix} \quad (9)$$

$$\underline{\mathbf{w}} = \begin{bmatrix} w[0] \\ \vdots \\ w[N-1] \end{bmatrix} \quad (10)$$

$$\underline{x} = \mathbf{S}\underline{\mathbf{a}} + \underline{\mathbf{w}} \quad (11)$$

Now the cross correlation function of $x[n]$ is calculate

$$\begin{aligned} \mathbf{R} &= E\{\underline{x}\underline{x}^H\} \\ &= E\{(\mathbf{S}\underline{\mathbf{a}} + \underline{\mathbf{w}})(\underline{\mathbf{a}}^H \mathbf{S}^H + \underline{\mathbf{w}}^H)\} \\ &= \mathbf{R}_s + \sigma^2 \mathbf{I}_N \end{aligned} \quad (12)$$

Where $\mathbf{I}_N$ is the unity (identity) matrix and $\mathbf{R}_s$ is equal to $\mathbf{S}\underline{\mathbf{a}}\underline{\mathbf{a}}^H \mathbf{S}^H$. Therefor the order of $\mathbf{R}_s$ matrix is equal to 2 while it has 2 independent column, based on these assumptions $\mathbf{R}_s$ matrix has two non-zero values of $\lambda_0, \lambda_1$ and its rest eigenvalues are zero.

**R** matrix has two eigen values as $\lambda_0 + \sigma$ and $\lambda_1 + \sigma$ and its rest values are equal to $\sigma$ . With the assumption of $\underline{\mathbf{q}}_0, \underline{\mathbf{q}}_1, \underline{\mathbf{q}}_2, \ldots, \underline{\mathbf{q}}_{N-1}$ as the eigen vectors of **R** matrix, the vectors from $\underline{\mathbf{q}}_2$ to $\underline{\mathbf{q}}_{N-1}$ are the former of the noise subspace. Assume vector $\underline{\mathbf{s}}(\omega)$ as:

$$\underline{\mathbf{s}}(\omega) = \begin{bmatrix} 1 \\ e^{j\omega} \\ \vdots \\ e^{j\omega(N-1)} \end{bmatrix} \quad (13)$$

Whereas signal and noise are orthogonal vectors, therefore $\underline{\mathbf{s}}(\omega)$ and vectors from $\underline{\mathbf{q}}_2$ to $\underline{\mathbf{q}}_{N-1}$ have to reaches their minimum values for the angular frequency estimation. So $P_{MUSIC}(\omega)$ is defined as

$$P_{MUSIC}(\omega) = \frac{1}{\sum_{m=2}^{N-1} |\underline{\mathbf{s}}^H(\omega)\underline{\mathbf{q}}_\mathbf{m}|^2} \quad (14)$$

The maximum points in $P_{MUSIC}$ are the desired points which their corresponding delay could be calculated by mathematical equations. Fig. 3 is truly the same as Fig. 2 that is designed by MUSIC algorithm. In this method, the direct message signal and the signal which received from antennas are sampled then their digital frequency will suppressed by DDC. Afterward, FFT and division operation calculated and then the signals cross through a band-pass filter. So far, the procedures of IFFT and MUSIC algorithms are the same. After that, by using MUSIC algorithm instead of IFFT, the target's delay is estimated. The results demonstrate the high range resolution of this method and its capability of close targets detection.

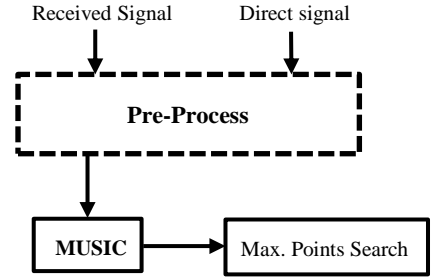

**Fig. 3** *Process box based on MUSIC algorithm*

*Simulations and Results:* In this section the range resolution of two close targets by using different number of FM channels with the aforementioned algorithms in section 3 and 4 has been simulated. The carrier frequency of FM signal is assumed to be 90 kHz which transfers to baseband frequency by using Mixer and DDC. The maximum frequency deviation ($f_\Delta$) and sampling frequency are 75 kHz and 2 MHz respectively.

Fig. 4 shows the real part of signal's IFFT of two targets with a 3 km distance which is passing from band-pass filter that is exactly the output of detection algorithm shown in Fig. 2 with one channel FM input signal. Fig. 5 demonstrates the delay time estimation of two targets with the distance of 2 km by applying MUSIC algorithm with one channel FM signal. As it's obvious in single channel FM broadcasting using MUSIC algorithm two targets with the minimum distance of 2 km are severable and distinguishable. Therefore, 1 km detection improvement over the IFFT algorithm obtained by MUSIC algorithm.

The real part of signal's IFFT, passing from band-pass filter for two targets with the distance of 2 and 1km by utilizing three and seven FM channels are shown in Fig. 6 and Fig. 8 respectively. The Delay time estimation of two targets with a distance of 1 km and 300 m using three and seven FM channels by MUSIC algorithm are shown in Figs. 7 and 9 respectively.

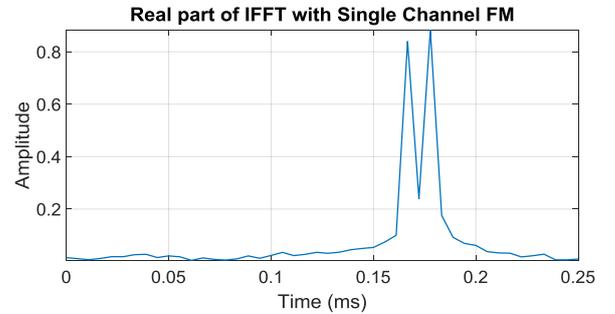

**Fig. 4** *The real part of the signal's IFFT which is passed from the band-pass filter for two targets with 3 k m distance*

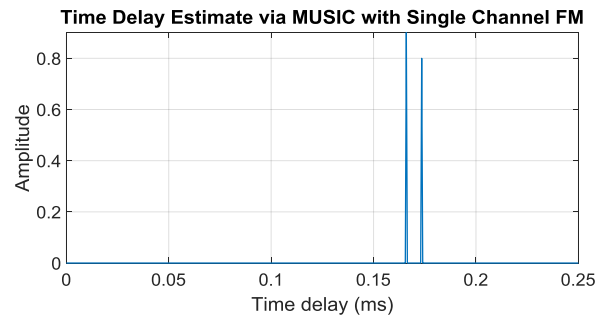

**Fig. 5** *Delay time estimation of two targets with the distance of 2 km by applying MUSIC algorithm with one channel FM*

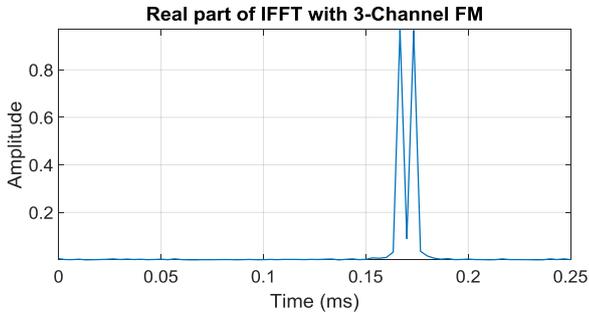

**Fig. 6** *The real part of the signal's IFFT which is passed from the band-pass filter for two targets with 2k m distance*

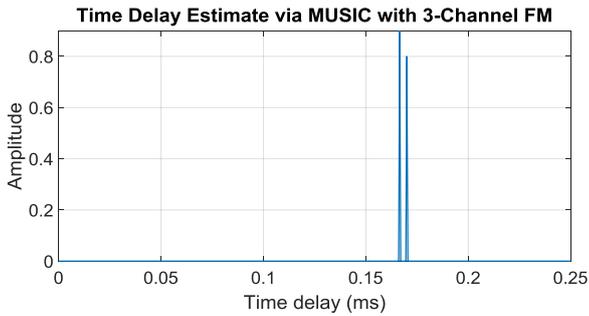

**Fig. 7** *Delay time estimation of two targets with the distance of 1 km by applying MUSIC algorithm with one channel FM*

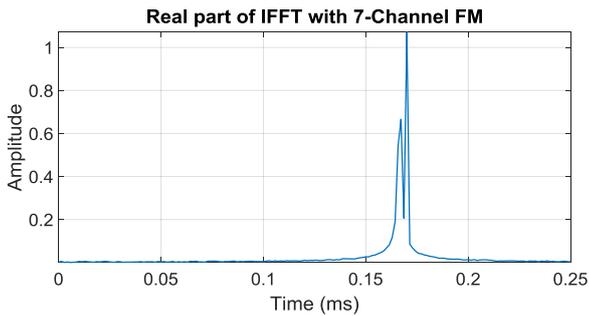

**Fig. 8** *The real part of the signal's IFFT which is passed from the band-pass filter for two targets with 1k m distance*

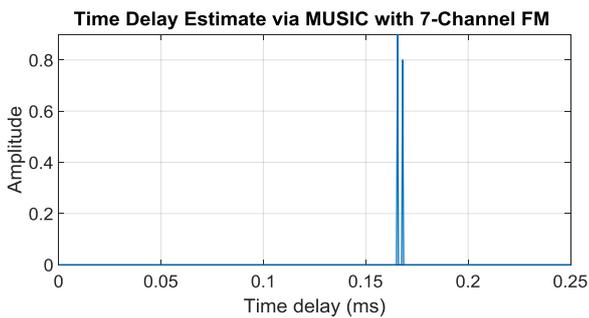

**Fig. 9** *Delay time estimation of two targets with the distance of 300m by applying MUSIC algorithm with one channel FM*

With the ever increasing in occupied FM channel, the range resolution of two targets improves. The obtained range resolution in MUSIC algorithm is twice as well as IFFT method. The comparison of range resolution between IFFT and MUSIC methods is depicted briefly in Table 1.

Also, Fig. 10 shows error percentage of IFFT and MUSIC methods by using different number of FM channels. The error of IFFT is obviously less than MUSIC algorithm. This error is obtained by getting average over 1000 iterations using Monte Carlo method.

**Table 1:** The comparison of range resolution between IFFT and MUSIC methods.

| Number of channels | IFFT | MUSIC | Improvement percentage |
|---|---|---|---|
| 1 | 3 km | 2 km | 150 % |
| 3 | 2 km | 1 km | 200 % |
| 7 | 1 km | 300 m | 300 % |

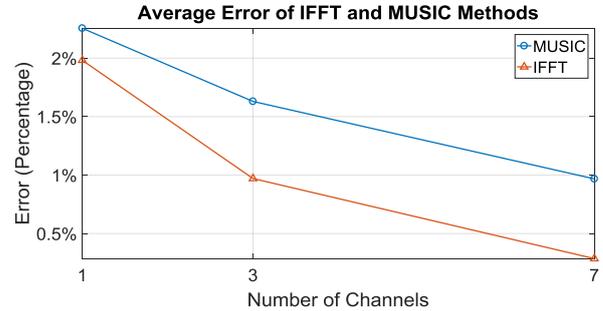

**Fig. 10** *Error percentage of IFFT and MUSIC methods*

*Conclusion*: In this paper the range resolution of two close targets is proposed by applying MUSIC and IFTT algorithms. In these methods, more FM channels have been occupied by transmitted signal to increase the bandwidth. Simulation results verified that the accuracy of the range resolution increased by using more FM channels and the close targets detects perfectly. The improvement of range resolution by using the same number of FM channel in MUSIC method is twice as good as IFFT but its Error is a little more than IFFT.

R. Ghavamirad, R.A. Sadeghzadeh, M.A. Sebt and A.H. Naderi (*Department of Electrical Engineering K. N. Toosi University of Technology, Tehran, Iran*)
E-mail: sadeghz@eetd.kntu.ac.ir